 \documentclass[twocolumn,prl,aps,psfig,showpacs,preprintnumbers,superscriptaddress]{revtex4}
\usepackage{graphicx}
\usepackage{epstopdf}
\usepackage{float}
\usepackage{color}
\usepackage{amsmath}

\newcommand{\red}[1]{\protect\textcolor{red}{#1}}

\begin{document}

\title{Hedgehog Spin-vortex Crystal Antiferromagnetic Quantum Criticality in CaK(Fe$_{1-x}$Ni$_x$)$_4$As$_4$ Revealed by NMR}
\author{Q.-P. Ding}
\affiliation{Ames Laboratory, U.S. DOE,  Ames, Iowa 50011, USA}
\affiliation{Department of Physics and Astronomy, Iowa State University, Ames, Iowa 50011, USA}
\author{W. R. Meier}
\affiliation{Ames Laboratory, U.S. DOE,  Ames, Iowa 50011, USA}
\affiliation{Department of Physics and Astronomy, Iowa State University, Ames, Iowa 50011, USA}
\author{J. Cui$\footnote[1]{Present address: Department of Chemistry, University of Delaware, Newark, Delaware 9716, USA}$}
\affiliation{Ames Laboratory, U.S. DOE,  Ames, Iowa 50011, USA}
\affiliation{Department of Chemistry, Iowa State University, Ames, Iowa 50011, USA}
\author{M. Xu}
\affiliation{Ames Laboratory, U.S. DOE,  Ames, Iowa 50011, USA}
\affiliation{Department of Physics and Astronomy, Iowa State University, Ames, Iowa 50011, USA}
\author{A. E. B\"ohmer$\footnote[2]{Present address: Karlsruhe Institute of Technology, Institut f\"{u}r Festk\"{o}rperphysik, 76021 Karlsruhe, Germany}$}
\affiliation{Ames Laboratory, U.S. DOE,  Ames, Iowa 50011, USA}
\affiliation{Department of Physics and Astronomy, Iowa State University, Ames, Iowa 50011, USA}
\author{S. L. Bud'ko}
\affiliation{Ames Laboratory, U.S. DOE,  Ames, Iowa 50011, USA}
\affiliation{Department of Physics and Astronomy, Iowa State University, Ames, Iowa 50011, USA}
\author{P. C. Canfield}
\affiliation{Ames Laboratory, U.S. DOE,  Ames, Iowa 50011, USA}
\affiliation{Department of Physics and Astronomy, Iowa State University, Ames, Iowa 50011, USA}
\author{Y. Furukawa}
\affiliation{Ames Laboratory, U.S. DOE,  Ames, Iowa 50011, USA}
\affiliation{Department of Physics and Astronomy, Iowa State University, Ames, Iowa 50011, USA}

\date{\today}

\begin{abstract} 

   Two ordering states, antiferromagnetism and nematicity, have been observed in most  iron-based superconductors (SCs).
 %  Both magnetic and nematic fluctuations have been proposed as the glue of Cooper pairs for superconductivity.
 % Therefore, which of them is the driving force, is a question of utmost importance, and is still under hot debate.
 % The antiferromagnetic (AFM) order in most iron-based SCs is closely accompanied by a nematic order.
   In contrast to those SCs,  the newly discovered SC CaK(Fe$_{1-x}$Ni$_x$)$_4$As$_4$ exhibits an antiferromagnetic (AFM) state, called hedgehog spin-vortex crystal structure,  without nematic order, providing the opportunity for the investigation into the relationship between spin fluctuations and SC without any effects of nematic fluctuations. 
 % We investigate the electronic and magnetic properties of CaK(Fe$_{1-x}$Ni$_x$)$_4$As$_4$ (0$\le x\le$ 0.049) using $^{75}$As NMR to search for the key to the pairing mechanism.
   Our $^{75}$As nuclear magnetic resonance studies on CaK(Fe$_{1-x}$Ni$_x$)$_4$As$_4$ (0$\le x\le$ 0.049) revealed that CaKFe$_4$As$_4$ is located close to a hidden hedgehog SVC AFM quantum-critical point (QCP).
   The magnetic QCP without nematicity in CaK(Fe$_{1-x}$Ni$_x$)$_4$As$_4$  highlights the close connection of spin fluctuations and superconductivity in iron-based SCs. 
   The advantage of stoichiometric composition also makes CaKFe$_4$As$_4$ an ideal platform for further detailed investigation of the relationship between magnetic QCP and superconductivity in iron-based SCs without disorder effects.

\end{abstract}

%  \pacs{74.70.Xa, 76.60.-k}
\maketitle

 % \section{Introduction} 
  %   Antiferromagnetism and nematicity, have been observed in iron-based superconductors (SCs).
    The relationship between antiferromagnetism (AFM), nematicity, and superconductivity in iron-based superconductors (SCs) has received wide interest \cite {Canfield2010,Johnston2010,Hirschfeld2011,Stewart2011}.
% which breaks the in-plane rotational symmetry while preserving time reversal symmetry, 
    The most controversial and important topic in iron-based SCs is the pairing mechanism, for which both spin and nematic fluctuations have been proposed, but no conclusion has been reached yet.
     SC in most iron-pnictides appears in proximity to an AFM ordered state,  which is closely accompanied by a nematic order.
    The SC critical temperature, $T_{\rm c}$, shows a dome-shaped dependence on carrier doping or pressure application. 
     Heavy fermions and high-$T_{\rm c}$ cuprates also exhibit similar phase diagrams \cite{Norman2005,Pfleiderer2009}.
      In these compounds, normal-state properties often show a striking deviation from Landau's Fermi liquid behavior near the optimal $T_{\rm c}$ composition \cite{Norman2005,Pfleiderer2009,Broun2008, Dai2009,Shibauchi2014}.
     The non-Fermi liquid behavior, such as temperature ($T$)-linear resistivity has been discussed in terms of a quantum critical point (QCP) \cite {Broun2008,Sachdev1999,Sachdev2011}. 
% which could be hidden inside the SC dome \cite {Broun2008,Sachdev1999,Sachdev2011}.
 %    What lies beneath this superconducting dome is a long-standing issue that remains unresolved \cite {Norman2005,Broun2008, Dai2009,Shibauchi2014}. 
 %    Whether a quantum critical point (QCP) is hidden inside it could be the key to understanding the unconventional superconductivity  \cite {Broun2008}. 
  %   QCP is the point where the phase transition occurs at absolute zero temperature driven by quantum fluctuations demanded by Heisenberg's uncertainty princible  \cite {Sachdev1999,Sachdev2011}.
   The existence of magnetic QCP has been reported in several iron-pnictide SCs, such as BaFe$_2$As$_2$-type (122) systems, for example, BaFe$_2$(As$_{1-x}$P$_x)_2$ \cite {Nakai2010, Shibauchi2014}, Ba(Fe$_{1-x}M_x$)$_2$As$_2$ ($M$ = Co, Ni) \cite{Ning2009,Ning2010,Zhou2013}.
 %   In the case of  BaFe$_2$(As$_{1-x}$P$_x)_2$, the existence of a QCP has been pointed out by the $T$-linear behavior of the resistivity, an increase in the effective mass from the de Haas-van Alphen effect, penetration depth measurements, as well as NMR measurements \cite{Shibauchi2014}.
%    For electron-doped Ba(Fe$_{1-x}$M$_x$)$_2$As$_2$ ($M$ = Co, Ni), the earlier NMR experiments report a magnetic QCP in these systems \cite{Ning2009,Ning2010,Zhou2013}. 
    However, recent NMR, $\mu$SR and neutron scattering measurements indicate the presence of inhomogeneous AFM domains near optimal SC, suggesting an avoided QCP in electron-doped Ba(Fe$_{1-x}M_x$)$_2$As$_2$ \cite {Dioguardi2013, Lu2014,Bernhard2012, Luo2012, Lu2013}.  
    The inhomogeneity presumably originates from the disorder due to Ni/Co substitutions.
%which makes it difficult to extract the intrinsic properties of  materials.
    Even though substitutional disorder is not introduced on the Fe plane in BaFe$_2$(As$_{1-x}$P$_x)_2$, a recent study combined with NMR, x-ray, and neutron diffraction measurements also suggests an avoided QCP \cite{Hu2015}.
% similar to the electron-doped 122 compounds. 
    Thus whether or not a magnetic QCP exists in 122 compounds is still under debate. 
%    Therefore, it is highly desired to find a new candidate to study a magnetic QCP without any disorder.  
     On the other hand, nematic fluctuations associated with nematic QCP have been increasingly suggested to enhance or even lead to SC in iron-based SCs, especially 122 systems \cite {Nem1,Nem2,Nem3}.

\begin{figure}[tb]
\includegraphics[width=\columnwidth]{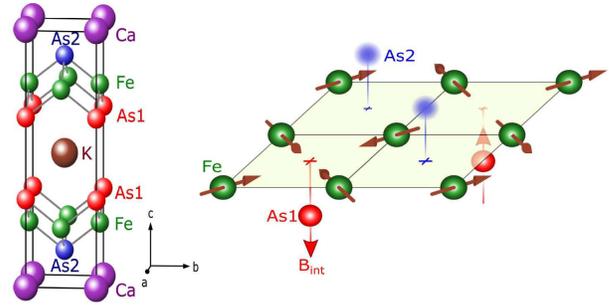} 
\caption{The crystal structure of CaKFe$_4$As$_4$ (left) and the sketch of hedgehog SVC spin structure on an Fe-As layer (right).
     The wine arrows represent the magnetic moments at the Fe sites and the red arrows represent the magnetic induction $B_{\rm int}$ at the As1 sites. 
}
\label{fig:Crystal}
\end{figure}

%    Here, we report that the newly discovered stoichiometric iron-based SC CaKFe$_4$As$_4$ \cite {Iyo2016} with $T_{\rm c} \sim$ 35 K \cite{Meier2016} can be a new candidate.
    Different from  the 122 family, the newly discovered stoichiometric iron-based SC CaKFe$_4$As$_4$ with $T_{\rm c} \sim$ 35 K adopts a structure where the Fe-As layers are separated by alternate Ca and K planes \cite {Iyo2016,Meier2016}.
    The segregation of Ca and K is driven by their dissimilar sizes and reduces the space group from  $I$4$/mmm$ in 122 compounds to $P$4$/mmm$ in CaKFe$_4$As$_4$.
 %    Recently, a new stoichiometric Fe-based SC, CaKFe$_4$As$_4$ (CaK1144), with $T_{\rm c} \sim$ 35 K has been discovered \cite {Iyo2016,Meier2016}. 
 %  CaKFe$_4$As$_4$ crystallizes through alternate stacking of the Ca and K layers across the Fe$_2$As$_2$ layer as a result of the large ionic radius difference \cite {Iyo2016,Meier2016}.
%   The ordering of the Ca and K layers changes the space group from $I$4$/mmm$ in $A$Fe$_2$As$_2$ to $P$4$/mmm$ in CaKFe$_4$As$_4$.
   Consequently, as shown in the left panel of Fig.\ \ref{fig:Crystal}, there are two inequivalent As sites: As1 and As2 sites close to the K and Ca layers, respectively.
   Other than SC, there is no indication of any other phase transition from 1.5 K to 300 K \cite {Meier2016, Xie2018}.
   The multiband nature of the compound and $s\pm$ nodeless two-gap SC state has been revealed by various techniques \cite{Jean2017,uSR2017,Cho2017,ARPES2016,Iida2017}.
% such as NMR \cite{Jean2017}, $\mu$SR \cite{uSR2017}, STM \cite{Cho2017}, ARPES \cite{ARPES2016}, and neutron scattering \cite{Iida2017} measurements. 
   Quite recently, a new magnetic state called hedgehog spin-vortex crystal (SVC) with tetragonal symmetry, as shown in the right panel of Fig.\ \ref{fig:Crystal}, has been discovered in the electron doped SC CaK(Fe$_{1-x}M_{x}$)$_4$As$_4$ ($M$ = Co or Ni) \cite{Meier20172}.
%    This is the first experimental observation of this AFM state which has been predicted by earlier theoretical studies \cite{Hoyer2016,Fernandes2016,O'Halloran2017}.    
    The magnetic phase transition has been shown to be second order \cite{Meier20172}.      
    The subsequent NMR and neutron diffraction studies clearly evidenced the microscopic coexistence of the hedgehog SVC and SC \cite {Ding2017,Kreyssig2018}.
    As the phase diagram of CaK(Fe$_{1-x}M_{x}$)$_4$As$_4$ is revealed to be similar to that of doped 122 systems \cite{Meier20172}, it is interesting to search for a QCP in the new system.
%    In the case of $M$ = Ni, $T_{\rm c}$ decreases from 35 K at $x$ = 0 to $\sim$ 10 K at $x$ = 0.049, and the new hedgehog SVC magnetic state appears by $x$ = 0.033 with a N\'eel temperature ($T_{\rm N}$) of 45 K which increases to 52 K at $x$ = 0.049 \cite{Meier20172}. 
%    Here, an interesting question is whether or not a magnetic QCP exists  in the new magnetic SC system.

\begin{figure}[tb]
\includegraphics[width=\columnwidth]{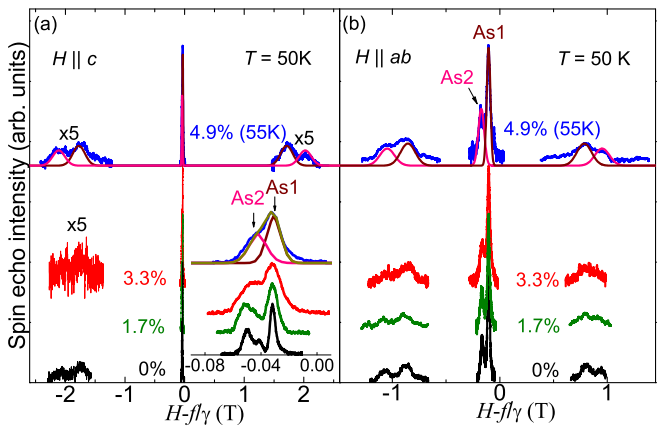} 
\caption{Field-swept $^{75}$As-NMR spectra of CaK(Fe$_{1-x}$Ni$_x$)$_4$As$_4$ at 50 K (55 K for 4.9$\%$ sample), for $H$ $\parallel$ $c$ axis (a) and $H$ $\parallel$ $ab$ plane (b). 
    The inset in (a) enlarges the central transition.
Pink and wine curves represent simulated spectra for As1 and As2, respectively. The dark yellow in the inset of (a) represents the sum of the simulated spectra. 
}
\label{fig:Spectra}
\end{figure}   

    In this Letter,  we show the experimental evidence which suggests that a hidden AFM QCP exists around $x$ = 0 from systematic $^{75}$As NMR measurements on CaK(Fe$_{1-x}$Ni$_x$)$_4$As$_4$ single crystals. 
    In particular, this is the first indication of a QCP with the newly discovered hedgehog SVC state.
    Nematic order and associated fluctuations, as well as the disorder caused by substitution, are absent in the stoichiometric SC CaKFe$_4$As$_4$ \cite{Raman}. 
% with the highest $T_{\rm c}$ 
    This makes CaKFe$_4$As$_4$ ideal for investigation into the potential link between spin fluctuations and superconductivity near the QCP without any effects from  the nematic fluctuations and/or disorder \cite{Unify}.
%   The absence of nematic order \cite{Meier20172} and nematic fluctuations \cite{Raman} in this system makes it completely different from 122 compounds and other well-studied iron-based SCs, the AFM order in the latter is always closely accompanied by a nematic order \cite{Unify}.
    The details of the NMR spectrum and nuclear spin-lattice relaxation rate 1/$T_1$ data in the hedgehog SVC magnetic state for $x$ = 0.049 and also in the SC state for $x$ = 0 have been reported previously \cite{Jean2017, Ding2017}. 
    In this paper, we mainly report the NMR data in the paramagnetic (PM) state and discuss the evolution of the hedgehog SVC spin correlations with Ni substitution.
% evidencing the AFM QCP around $x$ = 0. 

%   The ground state of the system can be tuned by either pressure or chemical doping \cite{hct2017,Meier20172}.

 %\section{Experimental}

   Single crystals of CaK(Fe$_{1-x}$Ni$_x$)$_4$As$_4$ for the NMR measurements were grown out of a high-temperature solution rich in transition-metals and arsenic \cite {Meier20172,Meier2016,Meier2017}. 
    Ni concentration was determined by WDS \cite {Meier20172}. 
    NMR measurements of $^{75}$As ($I$ = $\frac{3}{2}$, $\frac{\gamma_{\rm N}}{2\pi}$ = 7.2919 MHz/T, $Q=$ 0.29 barns) nuclei were conducted using a lab-built phase-coherent spin-echo pulse spectrometer on four different compounds, pure ($T_{\rm c}$ = 35 K) and 1.7$\%$ substitution ($T_{\rm c}$ =  31 K), 3.3$\%$ substitution ($T_{\rm c}$ = 23 K, $T_{\rm N}$ = 45 K), and 4.9$\%$ substitution ($T_{\rm c}$ = 10 K, $T_{\rm N}$ = 52 K) \cite{Meier20172}.  
%   In situ ac magnetic susceptibility ($\chi_{\rm ac}$)  was measured by monitoring the resonance frequency $f$ of the NMR coil tank circuit  as a function of temperature ($T$)  using a network analyzer.
   The $^{75}$As-NMR spectra were obtained by sweeping the magnetic field $H$ at fixed frequencies.
   The $^{75}$As 1/$T_{\rm 1}$ was measured with a saturation recovery method \cite{T1}.
%   The details NMR data in the hedgehog SVC magnetic state for $x$ = 0.049 and in the SC state for $x$= 0 have been reported previously. 
 %  In this paper, we mainly report the NMR data in the paramagnetic state for the four compounds and discuss the evolution of the hedgehog-type AFM spin correlations with Ni substitution.

%  \section{Results and discussion}

   Figures\ \ref{fig:Spectra} (a) and (b) show the typical field-swept $^{75}$As-NMR spectra of CaK(Fe$_{1-x}$Ni$_x$)$_4$As$_4$ in the PM state  for two magnetic field directions, $H$ $\parallel$ $c$ axis and $H$ $\parallel$ $ab$ plane, respectively. 
   As reported in the previous papers \cite{Jean2017, Meier20172, Ding2017}, the two sets of $I=3/2$ quadrupole split lines,  corresponding to the two inequivalent As sites, are observed for the four compounds  as shown  in Fig. \ref{fig:Spectra}.
   The lower field central peak with a greater Knight shift $K$ (and also larger quadrupole frequency, $\nu_{\rm Q}$)  has been assigned to the As2 site close the Ca layers and the higher field central peak with a smaller $K$ (and also smaller $\nu_{\rm Q}$) has been attributed to the As1 site close to the K layers \cite{Jean2017}.
   The clear separation of the two As NMR lines indicates that the well ordered K and Ca layers are not disturbed by Ni substitution.
% although the linewidths of both the central and satellite lines increase due to the disorder introduced by the substitution. 
 %  The linewidth of the central line for the $H$ $\parallel$ $c$ axis  is nearly independent of $T$ in the PM state for all the samples.

\begin{figure}[tb]
\includegraphics[width=\columnwidth]{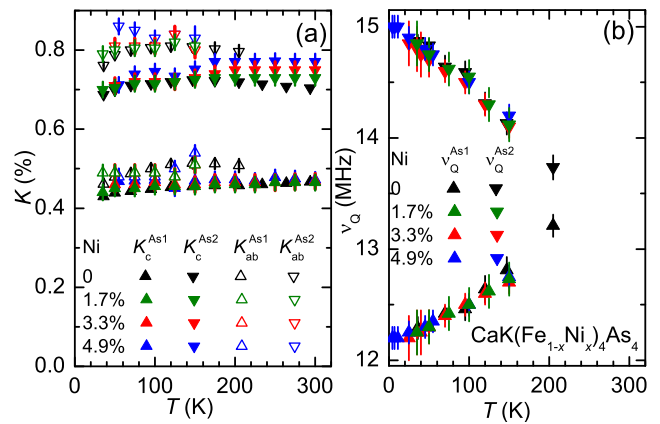} 
\caption{(a) $T$ dependences of the $^{75}$As-NMR shifts $K_c$ and $K_{ab}$ for the As1 and As2 sites. 
%The SC and AFM transition temperatures of the compounds are also listed.
(b) $T$ dependences of quadrupole frequency $\nu_{\rm Q}$ for the As1 and the As2 sites estimated from the NMR spectra.
}
\label{fig:K}
\end{figure}   

   Figures\ \ref{fig:K} (a) and (b)  show the $T$ dependence of $K_{ab}$ ($H$ $\parallel$ $ab$ plane), $K_c$ ($H$ $\parallel$ $c$ axis) and $\nu_{\rm Q}$ for the two As sites. 
  The As2 site $K$-values are uniformly higher than the As1 site $K$-values over the full temperature range.
   Due to the poor signal intensity at high $T$, $\nu_{\rm Q}$ and $K_{ab}$ can only be determined precisely up to 150--200 K. 
   $K_{c}$, though, can be determined up to 300 K.
   For both the As sites, the $K$-values are nearly independent of $T$, and also nearly independent of Ni substitution, indicating that static uniform magnetic susceptibility is nearly independent of both $T$ and $x$, although the ground states vary from magnetic to nonmagnetic. 
   These data also suggest that Ni substitution up to 4.9$\%$ does not produce significant change in the density of states at the Fermi energy $N(E_{\rm F})$.

% and the uniform spin susceptibility $\chi^{\prime}(\vec{q}, \omega_0)$ with $q$ = 0 and $\omega_0$ = 0. 
   
   The $T$ dependences of $\nu_{\rm Q}$ of As1 and As2 in CaK(Fe$_{1-x}$Ni$_x$)$_4$As$_4$ are shown in Fig.\ \ref{fig:K}(b).
   Similar to the $K$ data, $\nu_{\rm Q}$ is almost independent of Ni substitution. 
   All the As2 $\nu_{\rm Q}$-values are larger than the As1 $\nu_{\rm Q}$-values.    
 % These results suggest that all static physics properties in the paramagnetic state is nearly independent of $x$. 
    For all the samples, with increasing $T$, $\nu_{\rm Q}$ of As1 increases, while $\nu_{\rm Q}$  of As2 shows an opposite trend.
   The different $T$ dependences of the values of $\nu_{\rm Q}$ for the two As sites have been ascribed to hedgehog SVC magnetic fluctuations \cite{Jean2017}.
  %  The static physical properties in the PM state are nearly independent of $x$ as discussed above, but the dynamical properties of the compounds are largely affected by Ni substitution as shown  below.

\begin{figure}[tb]
\includegraphics[width=\columnwidth]{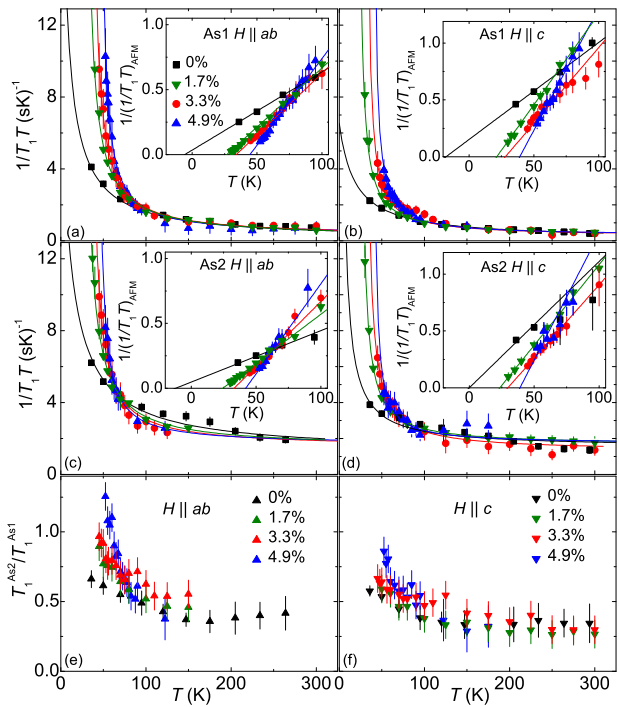} 
\caption{$T$ dependence of 1/$T_1T$ of As1 site for $H$ $\parallel$ $ab$ plane (a) and $H$ $\parallel$ $c$ axis (b), As2 site for $H$ $\parallel$ $ab$ plane (c) and $H$ $\parallel$ $c$ axis (d).
The solid lines represent fits to 1/$T_1T = C/(T+\theta)$ + constant (see text). 
Inset: Inverse of the temperature dependent component of 1/$T_1T$. The intercepts of the linear fits with the $x$ axis correspond to $-\theta$ (see text).
$T$ dependences of $T_{\rm 1}^{\rm As2}$/$T_{\rm 1}^{\rm As1}$ for $H$ $\parallel$ $ab$ plane (e), and $H$ $\parallel$ $c$ axis (f).
 }
\label{fig:T1}
\end{figure}

%   Low-energy AFM fluctuations are probed in all the samples via 1/$T_1$ measurement.
   Figures\ \ref{fig:T1} (a) - (d) show the $T$ dependence of $1/T_1T$ for the two As sites under the different magnetic field directions.
   As can be seen, all $1/T_1T$ plots increase with decreasing $T$.     
   $1/T_1T$ can be expressed in terms of the imaginary part of the dynamic susceptibility $\chi^{\prime\prime}(\vec{q}, \omega_0)$ at the Larmor frequency $\omega_0$ per mole of electronic spins as \cite{Moriya1963},
  $\frac{1}{T_1T}=\frac{2\gamma^{2}_{N}k_{\rm B}}{N_{\rm A}^{2}}\sum_{\vec{q}}|A(\vec{q})|^2\frac{\chi^{\prime\prime}(\vec{q}, \omega_0)}{\omega_0}$, 
   where the sum is over the wave vectors $\vec{q}$ within the first Brillouin zone, $A(\vec{q})$ is the form factor of the hyperfine interactions.
% and $\chi^{\prime\prime}(\vec{q}, \omega_0)$  is the imaginary part of the dynamic susceptibility at the Larmor frequency $\omega_0$. 
   On the other hand, $K$ reflects the $q$ = 0 component of magnetic susceptibility $\chi^{\prime}(\vec{q}, \omega_0)$.
   Therefore, the enhancement  of $1/T_{\rm 1}T$ at low $T$ where $K$ is almost constant  indicates  a growth of AFM spin correlations with $q$ $\neq$ 0. 
  
    According to the previous analysis of 1/$T_1T$ for $x$ = 0 and 0.049  \cite{Jean2017, Ding2017}, the AFM spin correlations can be characterized to be hedgehog SVC correlations experimentally by the $T$ dependence of  the ratio of 1/$T_{\rm 1}$ for As1 and 1/$T_{\rm 1}$ for As2 [Figs.\ \ref{fig:T1}(e) and (f)].
    In the hedgehog SVC ordered state, the internal magnetic induction $B_{\rm int}$  at As1 is finite along the $c$ axis while the $B_{\rm int}$  at As2 is zero due to a cancellation originating from the characteristic spin structure \cite{Meier20172,Ding2017}.  
     Therefore, one expects that 1/$T_1$ for As1 is enhanced compared to that of As2 if the AFM spin fluctuations originate from the hedgehog SVC-type spin correlations.  
     For stripe-type AFM fluctuations, the $T$ dependence of $1/T_1$ for As1 should scale to that of $1/T_1$ for As2 since there is no cancellation of the internal induction at either As site \cite{Meier20172}.  
     As can be seen in Figs.\ \ref{fig:T1}(e) and (f), the ratios of $T_1^{\rm As2}/T_1^{\rm As1}$ show a nearly $T$ independent value of $\sim$ 0.3--0.4 above 150 K,  which could be determined by the different hyperfine coupling constants for the As1 and As2 sites. 
     Below 150 K,  clear enhancements of $T_1^{\rm As2}/T_1^{\rm As1}$ are observed.
%     As shown in Fig.\ \ref{fig:T1}(e), the enhancement of ratio $r$ $\equiv$ $T_{1c}$/$T_{1ab}$ for As1 is also larger than that for As2 below 150 K.
     These results indicate that the As1 sites experience stronger AFM spin fluctuations than the As2 sites, consistent with the growth of the hedgehog SVC spin fluctuations.

      In order to get more insight into the hedgehog SVC spin fluctuations, we analyze the 1/$T_1T$ data using a  phenomenological model where $1/T_1T$ is decomposed into two components:  $1/T_1T$ = $(1/T_1T)_{\rm{AFM}}$ + $(1/T_1T)_{\rm{0}}$   \cite {Ning2010,Ning2009,Nakai2010,Zhou2013,Cui2015}.  
      Here, $(1/T_1T)_{\rm{AFM}}$ is due to the AFM spin fluctuations which can be described by  a Curie-Weiss (CW) formula  $(1/T_1T)_{\rm{AFM}} = C/(T+ \theta)$  expected for an AFM spin fluctuation for a two-dimensional (2D) system from the self-consistent renormalization theory \cite{Moriya1963}. 
      The Curie constant $C$ measures the spectral weight of AFM fluctuations and the Weiss temperature $\theta$ corresponds to the distance  from the AFM instability point.
      For the other term originating from a contribution other than the AFM correlations, we assumed $(1/T_1T)_{\rm{0}}$ = constant  which is expected for Korringa relation of $T_1TK^2$ = constant since $K$ is nearly constant.    
     The solid lines in Figs.\ \ref{fig:T1} (a)-(d) are fits for all 1/$T_1T$ data.   
      The CW behavior of (1/$T_1T)_{\rm AFM}$ can be also seen clearly in the insets for each figure where the inverse of (1/$T_1T)_{\rm AFM}$ follows $T$ linear dependence.

\begin{figure}[tb]
\includegraphics[width=\columnwidth]{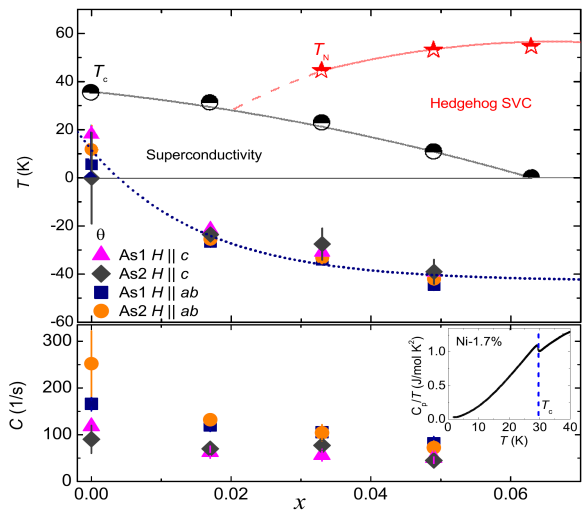} 
\caption{ Phase diagram of Ni doped CaKFe$_4$As$_4$ (Top). 
$T_{\rm N}$ and $T_{\rm c}$ are from Ref. \onlinecite {Meier20172}.
%$T_{\rm N}$ and $T_{\rm c}$ denote the AFM, and SC transtition temperature, respectively. 
Weiss temperature $\theta$ and Curie constant $C$ (Bottom) are obtained from the fits in Fig.\ \ref{fig:T1}. 
%The pressure evolution of phase diagram is from Ref. \onlinecite{hct2017}.
% Here, hcT indicates half-collapsed-tetragonal phase.
The lines are the guides to the eyes.
The inset of the bottom panel shows the specific heat $C_p/T$ of Ni-1.7\% sample.
}
\label{fig:Phase}
\end{figure}   
  
   Figure \ \ref{fig:Phase}~shows the $x$ dependence of $\theta$ and $C$ estimated from the fits, together with the phase diagram of CaK(Fe$_{1-x}$Ni$_{x}$)$_4$As$_4$ \cite{Meier20172}. 
      As shown in the bottom panel,   as $x$ decreases, $C$ increases and $\theta$ approaches zero. 
%    The magnetic fluctuations in PM state appear to head towards a putative QCP as $x$ $\to$ 0.
     The increase of $C$ suggests enhancement of the AFM fluctuations. % towards the highest $T_{\rm c}$. 
     A similar trend  of $T_{\rm c}$ is observed, suggesting a correlation between  $C$  and $T_{\rm c}$.   
    On the other hand, as shown above, $N(E_{\rm F})$ in CaK(Fe$_{1-x}$Ni$_{x}$)$_4$As$_4$ changes little upon $x$, while $T_{\rm c}$ decreases as $x$ increases. 
     This is in contrast to the conventional BCS superconductors, in which $N(E_{\rm F})$ generally correlates with $T_{\rm c}$.
   These results strongly, therefore, indicate that the hedgehog SVC spin fluctuations play an important role in the appearance of SC in CaKFe$_4$As$_4$. 
%      Furthermore, our results suggest the presence of an hedgehog-type AFM QCP near $x$ = 0. 
%    $|{\theta}|$ decreases with lowering Ni-substitution level and reaches nearly zero around $x$ = 0, where $T_{\rm c}$ reaches its highest value of $\sim$ 35 K. 
    $\theta$ = 0 K implies that the compound would show a hedgehog SVC magnetic order at 0 K \textit {if it remained in the normal state}.
    This is an indication of an AFM QCP around  $x$ = 0 which is hidden or, as shown below, shifted by the appearance of SC below $T_{\rm c}$.
% , suggesting a hidden AFM QCP.

    In general, one expects that a magnetic QCP exists in proximity to an AFM phase boundary. 
      For CaK(Fe$_{1-x}$Ni$_x$)$_4$As$_4$, therefore, an AFM state may be expected in the Ni-1.7\% sample whose $T_{\rm N}$ is expected to be lower than $T_{\rm c}$ from a simple extrapolation of the values of $T_{\rm N}$  of the higher Ni concentration compounds.
% as shown in the phase diagram. 
     As shown in  Figs. 3 (a)-(d), the strong divergent behavior in 1/$T_1T$ with the negative $\theta$ value observed in the $T > T_{\rm c}$ data for Ni-1.7\% sample is very similar to those observed at $T_{\rm N}$ for Ni-3.3\% and 4.9\% samples, suggesting an AFM order in the Ni-1.7\% doped sample.
%  based on the magnetic properties in the normal state. 
    The NMR spectrum measurements below $T_{\rm c}$ in Ni-1.7\% sample were difficult to determine the ground state due to the poor NMR signal intensity originating from Meissner effect.
    However, as shown in the inset of the bottom panel of Fig.\ \ref{fig:Phase}, specific-heat ($C_{\rm p}$) measurements  do not show any evidence of magnetic order below $T_{\rm c}$ in the Ni-1.7\% doped sample where the clear jump in $C_{\rm p}$ at $T_{\rm c}$  can be detected \cite{SpecificHeat}.  
%    In addition, we do not see any trace of the AFM order below $T_{\rm c}$ in Ni-1.7\% sample from NMR spectrum measurements, although the spectrum measurements were difficult due to the poor NMR signal intensity originating from Meissner effect.
    These results indicate that no static magnetic order is established in the Ni-1.7\% sample below $T_{\rm c}$ and in a similar manner there is no magnetic order expected for $x \leq$ 0.017.
%the magnetic order disappears below 1.7\% Ni substitution in the SC state.
% although the magnetic ordered state is suggested by the 1/$T_1T$ data above $T_{\rm c}$.    
% as schematically shown in the phase diagram in Fig.~\ \ref{fig:Phase}, 
     Similar sudden disappearance of the AFM state in SC state has been observed in 122 families where the $T_{\rm N}$ line bent backwards into the AFM region \cite{Johnston2010}.
    The absence of the magnetic order may originate from the strong competition with SC order in this region.
% which has been discussed in Ref. \onlinecite{Compet}.
    The disappearance of magnetic order and/or the backbending behavior of $T_{\rm N}$ have been discussed theoretically based on Ginzburg-Landau theory \cite {Compet}.
   %   It is interesting to point out that the AFM state disappears when  $T_{\rm N} < T_{\rm c}$, while coexistence of AFM and SC can be observed if  $T_{\rm N} > T_{\rm c}$. 
 %    Although the origin of the peculiar behavior of the  $T_{\rm N}$ line is still not well understood, a possible  qualitative explanation  could be as follows.
%     When an AFM state sets in, a part of the Fermi surface may contribute to  the AFM order state satisfying a nesting condition. 
%     But the system may have the residual Fermi surface which may produce a SC transition at lower temperature, making  the coexistence of AFM and SC possible. 
%     On the other hand, when  $T_{\rm N} < T_{\rm c}$, once SC state sets in, SC gap opens at all Fermi surface in $k$ space. 
%     This would make AFM transition difficult because of no residual Fermi surface satisfying a nesting condition for an AFM transition. 
%     If this were the case, the disappearance of AFM state may depend on SC gap symmetry such as S-wave or d-wave because of an existence of nodes. 
     Although the actual AFM phase boundary in SC state at low Ni-substitution levels does not change our conclusion of the hidden QCP around $x$ = 0, which is revealed from the 1/$T_1T$ data in the normal state, it is interesting and important to reveal  the details of the phase diagram of CaK(Fe$_{1-x}$Ni$_x$)$_4$As$_4$.  
    It would also be of great interest to investigate how the phase boundary changes if the SC is suppressed, for example, by strong magnetic field.

    It is generally believed that, in proximity to an AFM QCP, the scattering process of quasiparticles is strongly affected by the quantum fluctuations, leading to non-Fermi-liquid behavior.
   As a result, electrical resistivity ($\rho$) increases linearly with temperature, which has been actually observed in high-$T_{\rm c}$ cuprates and some iron-based SCs around a QCP \cite{Broun2008,Shibauchi2014,Sachdev2011,Zhou2013,Nakai2010}.
   In the case of CaKFe$_4$As$_4$ \cite{Meier2016}, on the other hand, $\rho$ is observed to be proportional to $T^{1.5}$ whose exponent is slightly higher than unity, but is clearly smaller than two expected for conventional Fermi liquid.
   Recently, the $T$-linear scattering rate of the coherent response \cite {Dai2013} as a result of quantum fluctuations  has been observed by an optical conductivity study of CaKFe$_4$As$_4$ \cite{Yang2017}, although the $T$-linear $\rho$ is not observed.  
   These results also support an existence of a magnetic QCP associated with the $T > T_{\rm c}$ normal state, around CaKFe$_4$As$_4$.

    The possibility of the avoided magnetic QCP in doped 122 compounds has been suggested by the observation of a distribution of  $T_1$ and also of a secondary incommensurate magnetic phase near the optimal $T_{\rm c}$ \cite {Dioguardi2013, Lu2014,Bernhard2012, Luo2012, Lu2013,Hu2015}.
    In the case of CaKFe$_4$As$_4$, it is important to point out that we do not observe any distribution of $T_1$ in the normal state, indicative of no significant inhomogeneity. 
    In addition, bulk measurements, $\mu$SR and neutron scattering as well as NMR also rule out any magnetic phase in CaKFe$_4$As$_4$ \cite{Meier2016,Xie2018,uSR2017,Jean2017}.
   Therefore the stoichiometric CaKFe$_4$As$_4$ provides a unique opportunity to explore quantum criticality without the inherent substitutional disorder in analogous doped 122 systems.
  $A$Fe$_2$As$_2$ ($A$ = K, Rb, Cs) \cite{Zhang2018,Wu2016} and LiFeAs \cite{Beak2013} are stoichiometric superconductors as well, and those systems are pointed out to be presumably located close to a magnetic instability.
  However, the maximum $T_{\rm c}$ in Ba$_{1-x}$K$_x$Fe$_2$As$_2$ appears at $x \sim$ 0.4, not at the stoichiometric composition.   
  LiFeAs shows a Fermi-liquid behavior, and no magnetic transitions has been detected in the temperature-doping ($T$-$x$) phase diagram of LiFe$_{1-x}$Co$_x$As \cite{Dai2015}.
   These are quite different from CaK(Fe$_{1-x}$Ni$_x$)$_4$As$_4$. 

     In summary, our $^{75}$As NMR measurements of CaK(Fe$_{1-x}$Ni$_x$)$_4$As$_4$ provide evidence for a QCP associated with hedgehog SVC fluctuations near $x$ = 0.
% existence of hedgehog SVC QCP at CaKFe$_4$As$_4$. 
    In contrast to other reported iron-based SCs with QCP, stoichiometric CaKFe$_4$As$_4$ is free of the substitutional disorder from chemical doping and also nematic transition.
%shows the highest $T_{\rm c}$, and it is stoichiometric and highly ordered. 
   The magnetic QCP without accompanying nematicity in CaK(Fe$_{1-x}$Ni$_x$)$_4$As$_4$  highlights the close connection of spin fluctuations and SC in iron-based SCs. 
    These advantages makes CaKFe$_4$As$_4$ an ideal platform to investigate the relationship between magnetic QCP and SC in iron-based SCs.

 %  \section{Acknowledgments} 
    We thank P. Wiecki for helpful discussions. 
    The research was supported by the U.S. Department of Energy (DOE), Office of Basic Energy Sciences, Division of Materials Sciences and Engineering. Ames Laboratory is operated for the U.S. DOE by Iowa State University under Contract No.~DE-AC02-07CH11358.
    W.R.M. was supported by the Gordon and Betty Moore Foundation's EPiQS Initiative through Grant No. GBMF4411.


\begin{thebibliography}{10}
\bibitem{SpecificHeat}  The heat capacity data of the Ni-1.7\% sample sample was measured using a hybrid adiabatic relaxation technique of the heat capacity option in a Quantum Design, Physical Properties Measurement System.
\end{thebibliography}
\end{document}